%% file: resopaper.tex
\documentclass[12pt,a4paper,dvips]{article}
\usepackage{a4p}
\usepackage{cite,mcite}
\usepackage{epsfig}
\usepackage{graphicx}
\usepackage{physics }
\usepackage{l3_title,ifthen}
\journalname{Phys. Lett. B}
\date{October 10, 2000}
\preprint{2000-129}
\newlength{\capindent}
\newlength{\capwidth}
\newlength{\figwidth}
\newcommand{\icaption}[2][!*!,!]{\hspace*{\capindent}%
  \begin{minipage}{\capwidth}
    \ifthenelse{\equal{#1}{!*!,!}}%
      {\caption{#2}}%
      {\caption[#1]{#2}}
  \end{minipage}}
\def\NP{Nucl. Phys. }
\def\PL{Phys. Lett. }
\def\ZfP{Z. Phys. }
\def\NIM{Nucl. Instr. Meth. }
\def\PRep{Phys. Rep. }
\def\PR{Phys. Rev. }
\def\PRL{Phys. Rev. Lett. }
\def\EURO{Eur. Phys. J. }

\def\gg{\ensuremath{\gamma \gamma}}
\def\Ggg{\ensuremath{\Gamma _{\gg}}}

\def\ra{\ensuremath{\rightarrow }}
\def\epem{\ensuremath{{\rm e}^+{\rm e}^-  }}
\def\pip{\ensuremath{\pi^+ }}
\def\pim{\ensuremath{\pi^- }}

\def\kos{\ensuremath{{\rm K}^{\rm 0}_{\rm S} }}
\def\k{\ensuremath{{\rm K}}}

\def\kskpi{\ensuremath{\kos\hspace{0.1em}\k^{\pm}\pi^{\mp}}}
\def\kkpi{\ensuremath{\k \bar{\k} \pi }}
\def\kstark{\ensuremath{\k ^*\hspace{0.1em}\k}}
\def\kstar{\ensuremath{\k ^*}}
\def\etapipi{\ensuremath{\eta \hspace{0.1em}\pi^+ \pi^- }}
\def\q2{\ensuremath{Q^2 }}
\def\e{\ensuremath{\eta\hspace{0.1em}{\rm (1440)}}}
\def\ea{\ensuremath{\eta\hspace{0.1em}{\rm (1295)}}}

\def\fb{\ensuremath{{\rm f}_1{\rm (1420)}}}
\def\fa{\ensuremath{{\rm f}_1{\rm (1285)}}}
\def\Spt{\ensuremath{(\sum \overrightarrow{p_T})^2}}
\def\spt{\ensuremath{P_T^2}}

\def\kkbar{\ensuremath{{\k}\bar {\k}}}
\def\BR{\ensuremath{\rm BR}}
\begin{document}

\begin{titlepage}
\title{
Light Resonances in  \boldmath{\kskpi}
and \boldmath{\etapipi} final states in \boldmath{\gg} collisions at LEP
}
\author{The L3 Collaboration}

\begin{abstract}
The $\ee \ra \ee \kskpi$ and $\ee \ra \ee \etapipi$  final states
are studied with the L3 detector at LEP using data collected
at centre--of--mass energies from 183~\GeV{} up to 202~\GeV{}. 
The mass spectrum  of the \kskpi{} final state shows an enhancement  around 1470~\MeV{}, 
which is identified with the pseudoscalar meson \e{}.
This state is observed in \gg{} collisions for the first time and
its two--photon width is measured to be 
$\Ggg\Bigl(\e\Bigr)\times \BR\Bigl(\e\ra{\rm K}\bar{\rm K}\pi\Bigr)={\rm 212}\pm{\rm 50}\stat\pm{\rm 23}\sys$~\eV{}.
Clear evidence is also obtained for the formation of the axial vector mesons f$_1$(1420) and f$_1$(1285).
In the \etapipi{} channel the \fa{} is observed, and upper limits for the 
formation of \e{} and \ea{} are obtained.
\end{abstract}

\vfill

\submitted
\end{titlepage}

\section{Introduction}

\label{intro}
Resonance formation in two--photon interactions offers a clean environment to study
the spectrum of mesonic states.
In this paper we study the reactions $\gamma\gamma\ra\kskpi$ and $\gamma\gamma\ra\etapipi$.  

The mass region between 1200 \MeV{} and 1500 \MeV{} is expected to contain several states \cite{PDG}.
For the pseudoscalar sector ($J^{PC}=0^{-+}$), the \e{} meson, formerly known as E or $\iota$, is expected to be seen. 
The \e{} was observed in hadron collisions and in radiative ${\rm J}$ decay, but not in two--photon collisions,
and only upper limits of its two--photon width, \Ggg{}, of the order of 1 \keV{} exist \cite{crystal,Cello}.
Therefore the \e{} may be interpreted as a prominent glueball candidate due to its 
strong production in a gluon rich environment.
There are, possibly, two pseudoscalars \cite{EXPS} in the 1440~\MeV{} mass region:
one at lower mass, $\eta_L$,  which decays into  a$_0 \pi$ or directly into $\eta \pi \pi$, 
and another at higher mass, $\eta _H$, decaying to $\kstark$.
The average masses and widths for these states \cite{PDG} are listed in Table~\ref{tab:etapar}.
Taking into account also the $\eta (1295)$, there are therefore three candidates for the
first radial excitations of the pseudoscalar SU(3) nonet. If one of these states is a gluonium,
its two--photon width is expected to be small with respect to a \qqbar{} state.
However, the gluonium  interpretation is disfavoured by lattice gauge theories \cite{mornings} 
which predict the lowest lying $0^{-+}$ gluonium state to be above 2~\GeV{}.
More  exotic interpretations such as a bound state of gluinos \cite{campos} are also proposed. 

Axial vector mesons ($J^{PC}=1^{++}$) are also present in these final states in the 1440~\MeV{} mass region.
The $\fb$ was observed in two--photon collisions, in the $\kkpi$ decay channel, 
by the CELLO \cite{Cello}, TPC--2$\gamma$ \cite{TPC}, JADE \cite{Jade} and Mark~II \cite{MRK2} Collaborations,
the \fa{} was seen \cite{MRK2,TPC} in the \etapipi{} decay channel.

\begin{figure}[p]
\begin{center}
\includegraphics[width=\figwidth ]{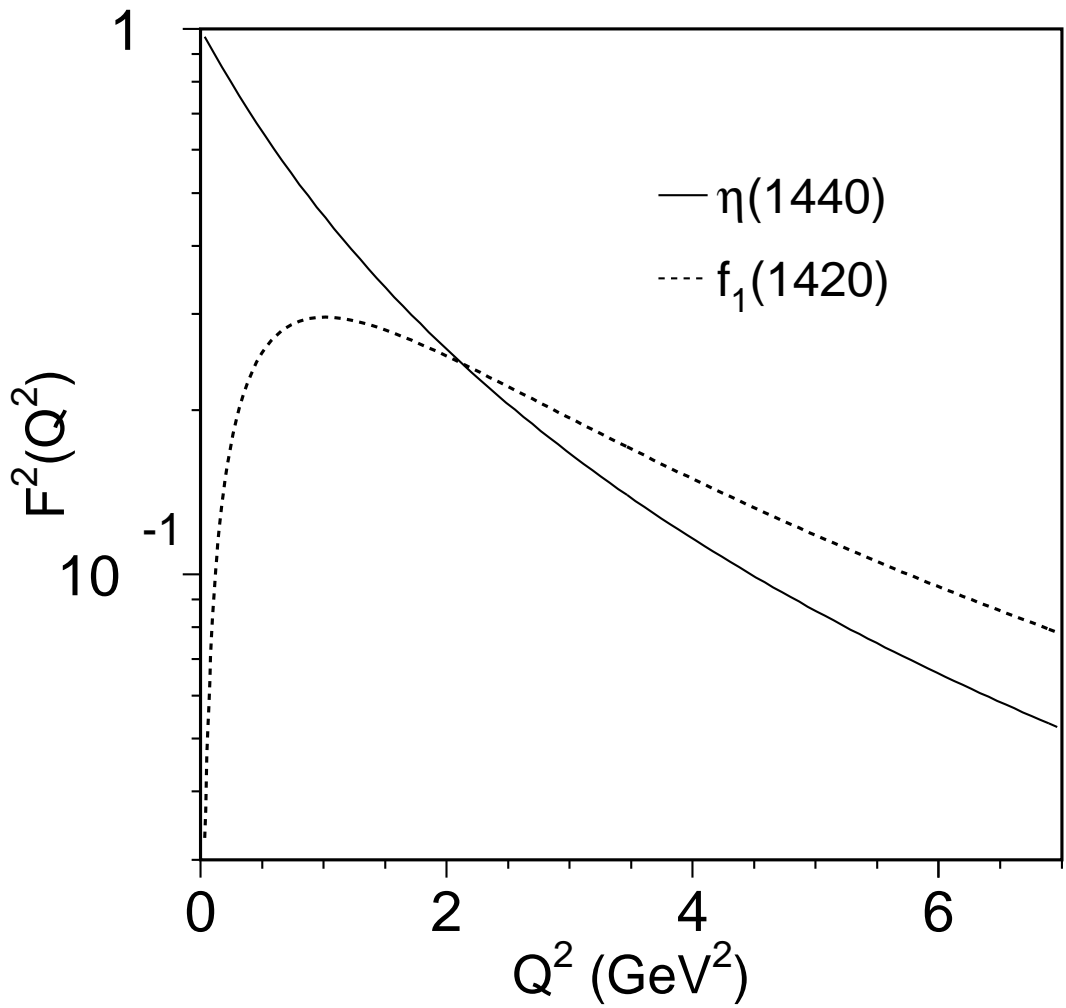}
\end{center}
\caption{ The form factors squared for \e{} (solid line) and \fb{} (dashed line)
in two--photon formation.}
\label{fig:formfactor}
\end{figure}

The data used here were collected by the L3 detector \cite{l3_000,*l3det:acc,*l3det:che,*l3det:ada,*l3det:bro}
at LEP from 1997 to 1999 at centre--of--mass energies between 183~\GeV{} and 202~\GeV{},
corresponding to a total integrated  luminosity of 449 pb$^{-1}$.
These events were collected by the track triggers \cite{LVT,*ITEC}.  
The analysis makes use of the dependence of the signal yield on the total transverse momentum $\spt = \Spt$, 
where the sum runs over all the observed particles. To a good approximation $\spt=\q2$, 
where \q2{} is the maximum virtuality of the two photons. 
According to the Landau--Yang theorem \cite{landau}, for real photons ($\q2\simeq 0$),
the production of a spin$-$0 state is allowed while that of a spin$-$1 state is suppressed.
In contrast, for virtual photons ($\q2 > 0$), the production of a spin$-$1 state is allowed while 
that of a spin$-$0 state diminishes. 
Such behaviour is described by the effective form factors for the pseudoscalar and axial vector mesons.
The form factors as a function of \q2{} calculated \cite{Schuler} for the \e{} and for the \fb{} 
are shown in Figure~\ref{fig:formfactor}. 
The difference at $\q2\simeq 0$ is clearly seen.
 
\section{Monte Carlo}
Two Monte Carlo generators are used to describe two--photon resonance formation:
EGPC \cite{Linde} and GaGaRes \cite{Gulik}.
The EGPC Monte Carlo describes the two--photon process as the product of the 
luminosity function for transverse photons \cite{Budnev}
and the resonance production cross section.
The latter is generated according to a Breit--Wigner function with $\Ggg = 1\keV$.
The ratio of the measured cross section to that obtained with the Monte Carlo
integration then gives $\Ggg$. The decay of the resonance is
generated according to Lorentz invariant phase space.

Events are generated for all the resonances of interest. 
For  the $\kskpi$ channel this is done for masses of  
1410~\MeV{}, 1440~\MeV{} and 1480~\MeV{} and a full width of 50~\MeV{}. 
For the $\etapipi$ channel, resonances are generated for the mass 1285~\MeV{} with a full width of 20~\MeV{} 
as well as for masses of 1405~\MeV{} and 1440~\MeV{} with 50~\MeV{} full width.

To take into account the \q2{} dependence of pseudoscalar and axial vector mesons,
EGPC events are re--weighted according to the prediction of GaGaRes. 
This uses the exact matrix element for resonance production, $\ee \ra \ee {\rm R}$,
obtained from the hard scattering approach \cite{Schuler}. 

The generated events are passed through the L3 detector simulation
based on the GEANT \cite{GEANT} and GEISHA \cite{GEISHA} programs.
Time dependent detector inefficiencies, as monitored during the data taking period,
are also simulated.

\section{The \boldmath{\epem\ra \epem \kskpi} channel}

\subsection{ Event selection}

\label{sec:kkpisel}
Events are selected by requiring four charged particles in the central tracker,
two tracks ($\k ^{\pm}\pi ^{\mp}$), coming from the interaction point,
and a $\kos$  decaying into $\pi ^+\pi ^-$ at a secondary vertex.
For charged particle identification, consistency of the corresponding $dE/dx$ measurement, 
with a confidence level CL$>$1\%\footnote{A $\chi^2_{dE/dx}$ is evaluated for each particle identity,
$\chi^2_{dE/dx} = (dE/dx_{measured} - dE/dx_{expected})^2\,/\,\sigma^2_{dE/dx}$}, 
is required.

Events with photons are rejected where
a photon is defined as an energy deposition in the electromagnetic calorimeter of more than 100~\MeV{}
in the polar angular range $0.21 < \theta < 2.93$ rad.
No track should lie within 0.2~rad from the photon direction. 

\smallskip
The $\kos\ra\pip\pim$ is identified by:
\begin{itemize}
\item\nointerlineskip
  a secondary vertex at least 3~mm away, in the transverse plane, from the interaction point;
\item\nointerlineskip
  the angle $\alpha$, in the transverse plane, between the flight direction of the $\kos$ candidate
  and the direction of the total transverse momentum vector of the two outgoing tracks being smaller than 0.2 rad;
\item\nointerlineskip 
  the momentum of the \kos{} candidate and the distance between the primary and the secondary vertex
  being consistent with the $\kos$ lifetime with CL$>$99\%;
\item\nointerlineskip
  the effective mass of the two pions being within the mass region 0.46--0.53~\GeV{}.
\end{itemize}\nointerlineskip

\begin{figure}[p]
\begin{center}
\includegraphics[width=\figwidth ]{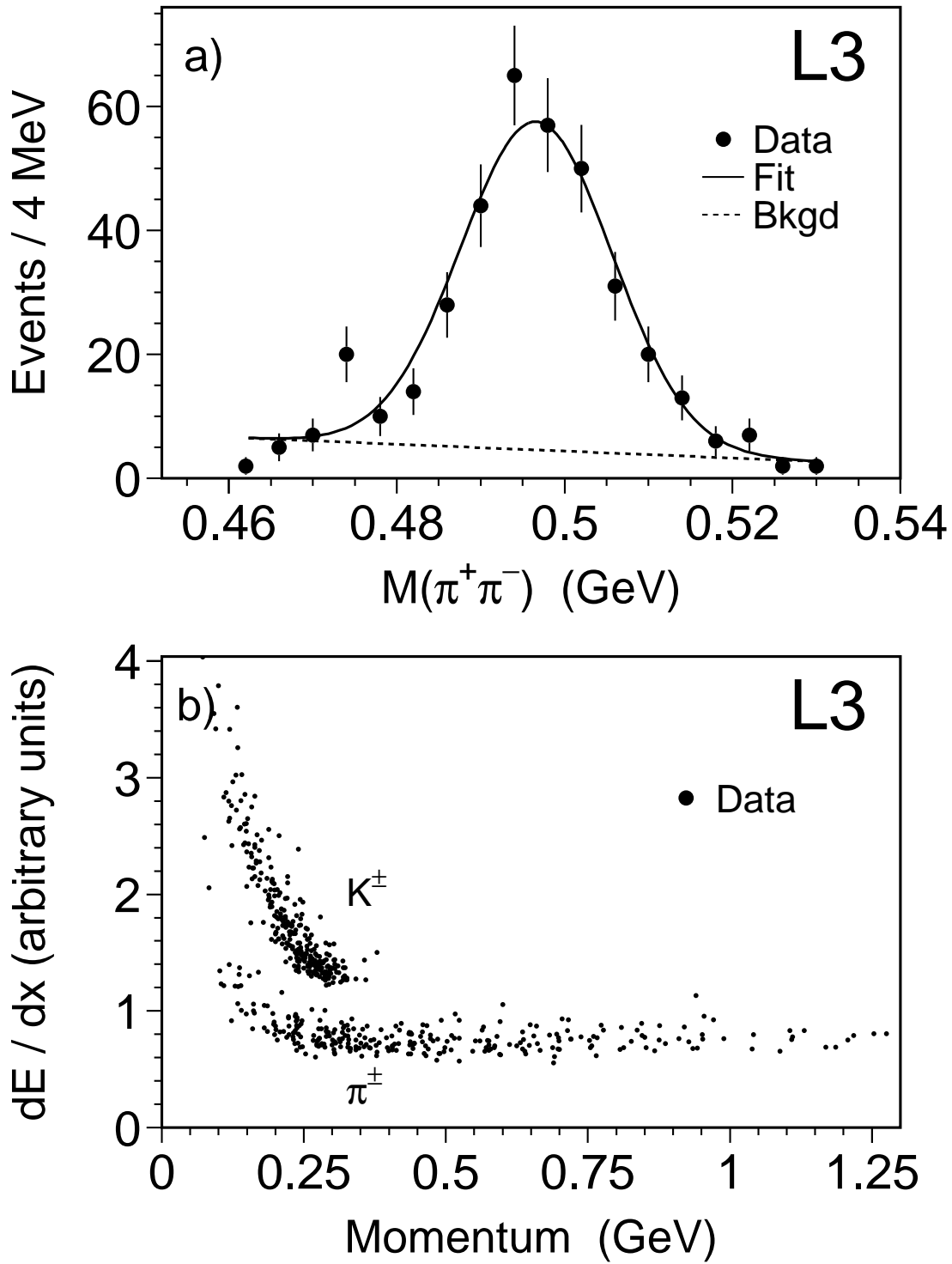}
\end{center}
\caption{ The \kskpi{} selection for $\spt < 0.2 \GeV^2$.
a) The \kos{} mass spectrum as the invariant mass of $\pip\pim$ coming from the secondary vertex.
b) $dE/dx$ of charged kaons and pions at the primary vertex.}
\label{fig:cuts}
\end{figure}

\noindent
$\kos \kos$ background, arising from $\k$/$\pi$ misidentification, is excluded
by rejecting events with a second $\kos$ in the mass range 0.48--0.51~\GeV{}.
Figure~\ref{fig:cuts}a presents the $\pip\pim$ invariant mass spectrum for particles from the secondary vertex
in events selected with $\spt < 0.2 \GeV^2$.
A Gaussian fit of the peak gives $M = 496.7 \pm 0.6 \MeV$ and $\sigma = 9.0\pm0.7\MeV$, 
consistent with the \kos{} mass and the detector resolution.

\smallskip
The $\k ^{\pm}\pi ^{\mp}$ identification requires that:
\begin{itemize}
\item\nointerlineskip 
  the two tracks lie within three standard deviations from the interaction point
  both in the transverse plane  and along the beam axis;
\item\nointerlineskip   
  their $dE/dx$ measurement must be consistent with a $\k ^{\pm}\pi ^{\mp}$ hypothesis,
  with the combined confidence level CL($\k^{\pm} \pi^{\mp}$)~$>$~1\%. 
  A pion is identified if  $\chi ^2_{dE/dx}(\pi^{\pm})<\chi ^2_{dE/dx}(\k^{\pm})-4$ and a kaon
  if $\chi ^2_{dE/dx}(\k^{\pm})<\chi ^2_{dE/dx}(\pi^{\pm})-4$.
  If the particles are not identified, both hypotheses are taken, each with a weight of 0.5.
\end{itemize}\nointerlineskip

To further improve the \kskpi{} selection, additional requirements are used as a function of \spt{}.
For low \spt{} the  $dE/dx$ identification has high discriminating power as shown in Figure~\ref{fig:cuts}b. 
For higher \spt{} the $dE/dx$ performance degrades, but the $\pip\pim\pip\pim$ background is smaller.
The data are hence divided into two samples.

\begin{itemize}
\item\nointerlineskip 
  For $\spt < 0.2 \GeV ^2$, the event is accepted if for the two tracks coming
  from  the primary vertex, 
  the two--pion hypothesis has a low confidence level, CL$(\pi^{\pm} \pi^{\mp}) < 10^{-3}$,
  from the $dE/dx$ measurement.
\item\nointerlineskip 
  For $\spt > 0.2 \GeV ^2$, either the previous requirement is satisfied or 
  two long tracks are reconstructed at the primary vertex and 
  the two tracks at the secondary vertex have a probability less than 30\% to come from the primary vertex.
\end{itemize}\nointerlineskip

\subsection{\boldmath{\q2} dependence}

\begin{figure}[p]
\begin{center}
\includegraphics[width=\figwidth ]{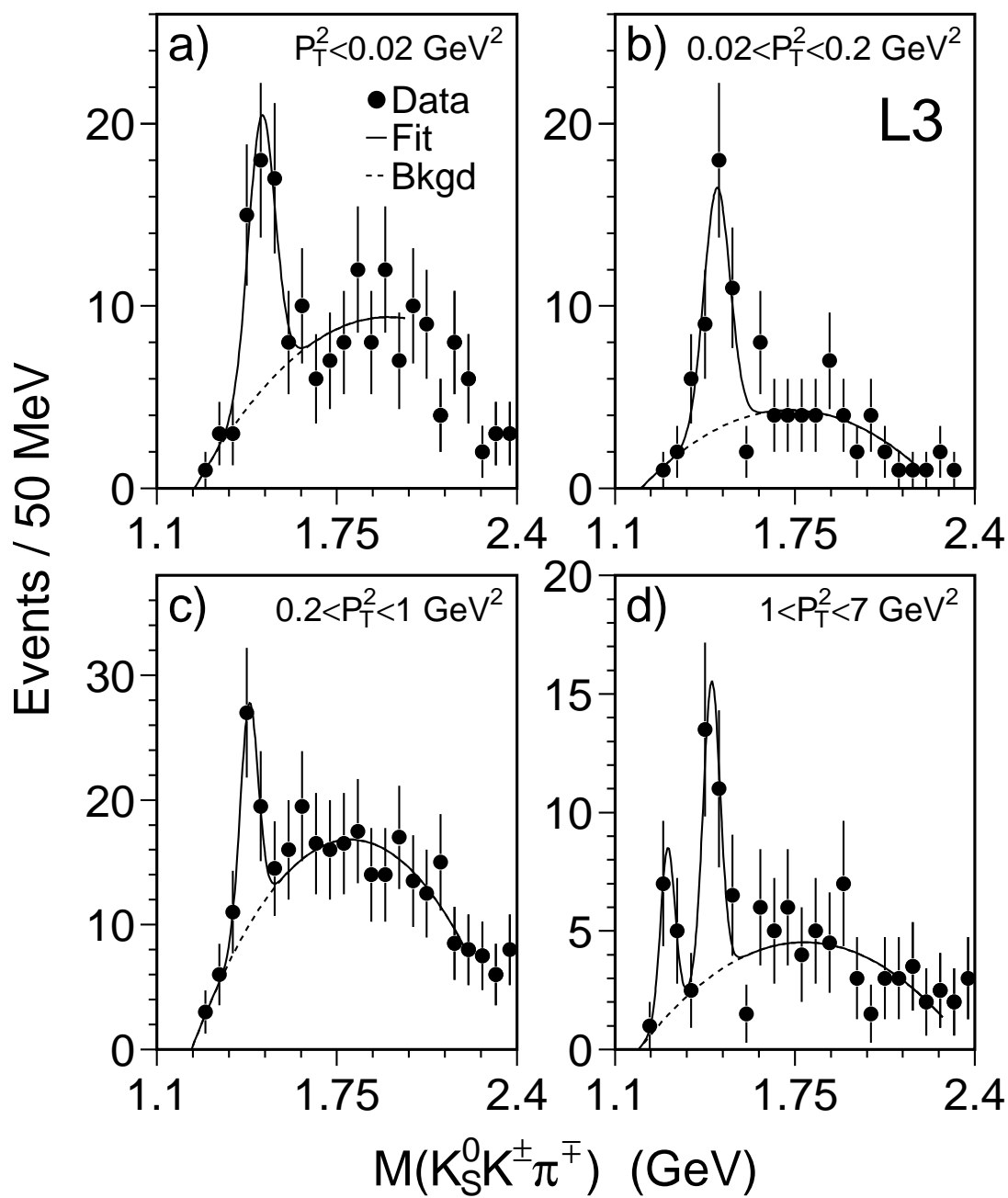}
\end{center}
\caption{ \kskpi{} spectra for different $\spt$ bins. 
The fits of a Gaussian plus polynomial background are superimposed on the data.
In the   highest $\spt$ bin, also the peak of the \fa{} is present
and the fit includes two Gaussians.}
\label{fig:ptbins}
\end{figure}

The selection results in the \kskpi{} effective mass spectra shown in Figure~\ref{fig:ptbins}.
A clear peak is seen in the 1440 \MeV {} mass region.
The data are subdivided into four \spt{} intervals of similar statistics 
in the peak region, in order to study the $\e$ and $\fb$ contributions.

A fit to a Gaussian plus second order polynomial background  is performed for each \spt{} interval.
The results are listed in Table~\ref{tab:mass}. 
For the peak in the lowest \spt{} interval, Figure~\ref{fig:ptbins}a, 
the mass and width, $M=1481\pm12\MeV$ and $\sigma=48\pm9\MeV$,
are compatible\footnote{A 
Gaussian fit to a Breit--Wigner of $\Gamma = 80\MeV$, convoluted with the resolution function 
$\sigma = 20\MeV$, gives $\sigma = 47\MeV$.} with those  of the $\eta _H\,$.
In the highest \spt{} interval, $1 \GeV ^2 <$ \spt{} $< 7 \GeV ^2 $,
an $\fa$ signal is present, so this spectrum is fitted to two Gaussians plus a polynomial.

The total efficiencies from the acceptance of the detector, the selection and trigger efficiencies, are presented in 
Table~\ref{tab:mass} for each
\spt{} interval. 
The detector acceptance and the selection efficiency,
as well as the efficiency of the central track trigger, 
which varies from 0.97 at low \spt{} to 0.74 at the highest \spt{}, are determined by Monte Carlo. 
Inefficiencies due to the higher level triggers, evaluated directly from the data, range from 2\% to 8\%.

\begin{figure}[p]
\begin{center}
\includegraphics[width=\figwidth ]{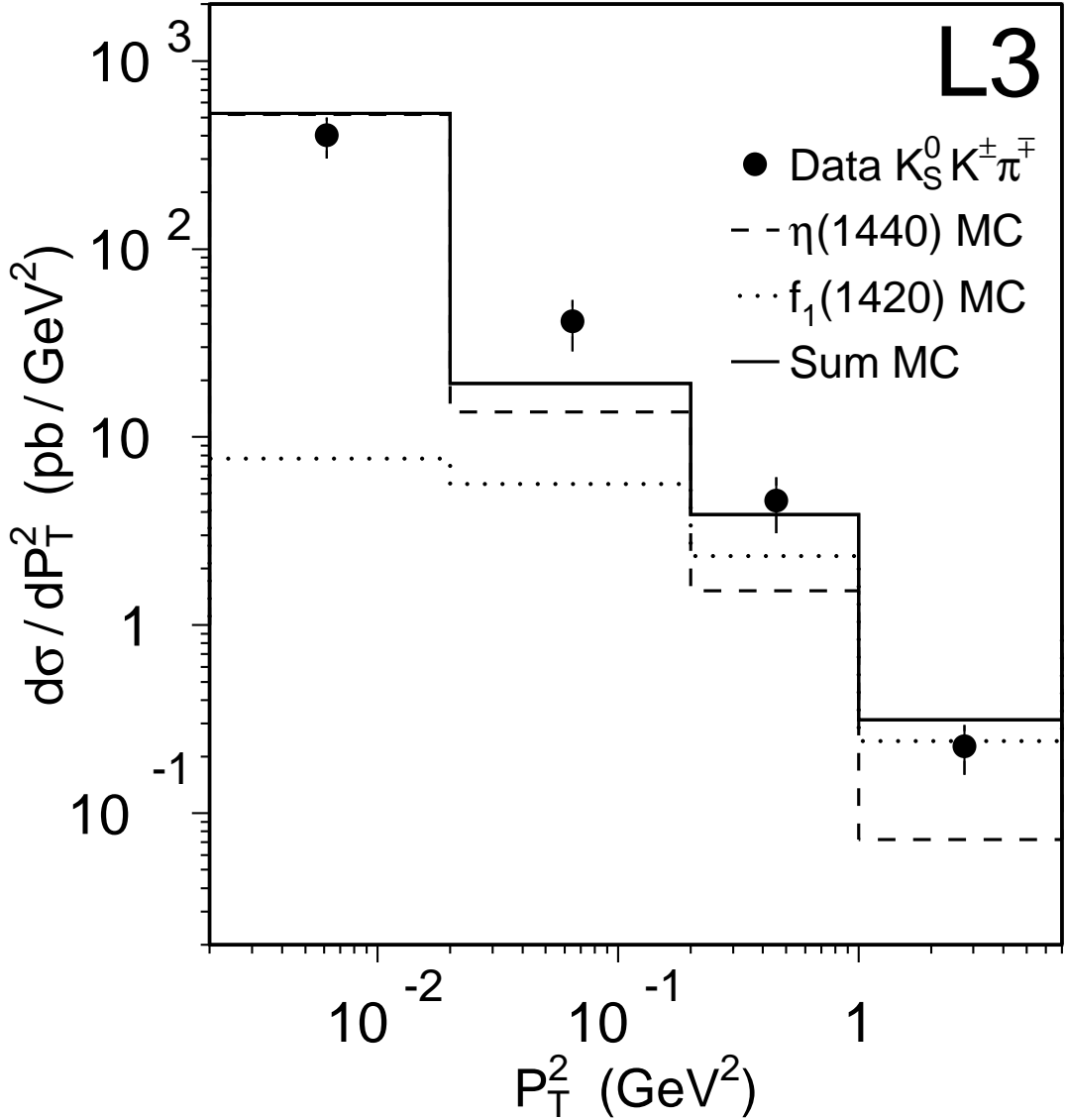}
\end{center}
\caption{ Differential cross section for the 1440~\MeV{} mass region, as estimated from the Gaussian fit, 
as a function of \spt{} in the \kskpi{} channel. 
The solid line is the sum of the \e{} and \fb{} simulations fitted to the data. 
The partial contributions of \e{} (dashed line) and \fb{} (dotted line) are also shown.}
\label{fig:dsdpt}
\end{figure}

 The cross section $\Delta \sigma $ for each \spt{} interval, evaluated from the fitted
 number of events in the Gaussian peak, is listed in Table~\ref{tab:mass}.
The differential cross section  $d\sigma /d \spt$ is shown in Figure~\ref{fig:dsdpt}.
It is fitted, using the predictions for the \q2{} dependence of  \e{} and \fb{} production  given by the GaGaRes Monte Carlo,
to the following three hypotheses.
\begin{itemize} 
\item\nointerlineskip 
  Pure pseudoscalar \e{}. This is excluded with a CL$\,\sim10^{-3}$, 
  due to the high number of events observed at high \q2{}.
\item\nointerlineskip 
  Pure axial vector meson \fb{}. This is  excluded with a CL$\,\sim10^{-5}$, 
  due to the strong peak at $\q2 \simeq 0$.
\item\nointerlineskip 
  Simultaneous presence of the \e{} and \fb{} resonances. 
  This hypothesis leads to a CL~$\simeq$~9\%, 
  with $74 \pm 10$ events for the spin--0 particle and $42 \pm 10$ events for the spin--1. 
\end{itemize}\nointerlineskip 

\subsection{Two--photon width}

The two--photon width of the $\e$ is determined using only the events with 
$\spt < 0.02 \GeV ^2$, shown in  Figure~\ref{fig:ptbins}a. 
This cut selects events produced by quasi--real photons, dominated by the  spin--0 state. 
The cross section measurement yields:
\begin{displaymath}
\Ggg\Bigl(\e\Bigr)\times \BR\Bigl(\e\ra\kos(\ra\pi ^+\pi ^-){\rm K}^{\pm}\pi ^{\mp}\Bigr)={\rm 49}\pm 12\stat  \eV{}. 
\end{displaymath}
\par
The most important systematic uncertainties result from varying
the $dE/dx$ cut (5\%), the \spt{} cut (4.7\%), and 
from the background subtraction and the shape of the resonance (8\%).
Using the result of the fit to the \spt{} dependence,  
the contribution of the $\fb$ is estimated to be  $1.5$\%.
This contribution is not corrected for, but included in the systematic uncertainty.

Taking into account the branching ratio values\cite{PDG}, $\BR(\kos\ra\pip\pim)$ and $\BR(\k^0\ra\kos)$, 
and the isospin factor $\k{\bar\k}\pi\,/\,\k^0\k^{\pm}\pi^{\mp}$
the two--photon width for the $\k{\bar \k}\pi$ decay channel is:
\begin{displaymath}
\Ggg\Bigl(\e\Bigr)\times \BR\Bigl(\e\ra{\rm K}\bar{\rm K}\pi\Bigr)={\rm 212}\pm{\rm 50}\stat \pm{\rm 23}\sys \eV{}.
\end{displaymath}
This value is consistent with the upper limit of 1.2 \keV{} reported by the CELLO Collaboration \cite{Cello}.

\subsection{Intermediate states}
\begin{figure}[p]
\begin{center}
\includegraphics[width=\figwidth ]{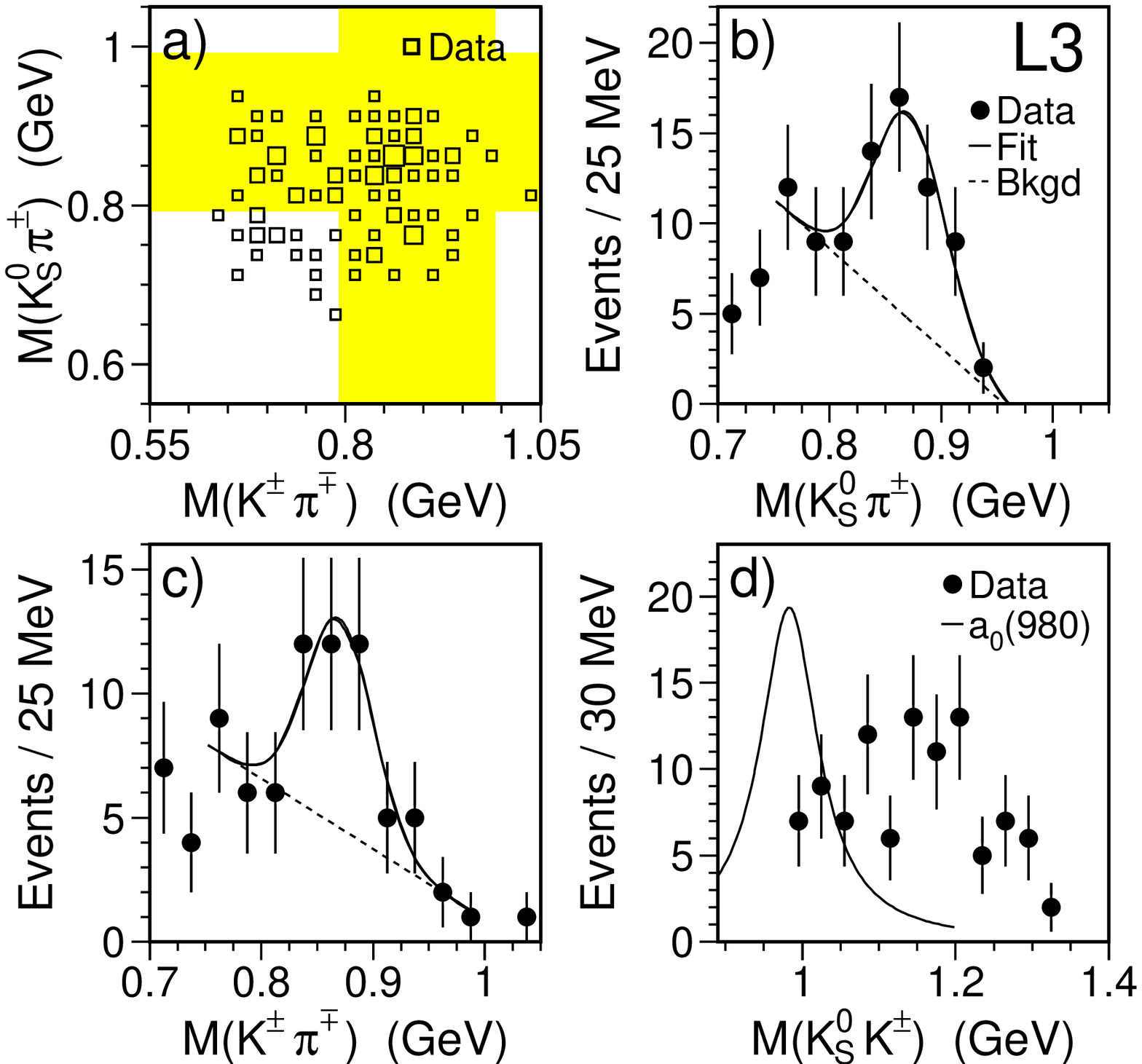}
\end{center}
\caption{ Search for the $\kstar(892)\k$ and a$_0(980)\pi$ states in the \kskpi{} 
sample for $\spt<0.2\GeV^2$ in the \e{} region (1370--1560~\MeV{}). 
The shadowed area displays the $\kstar(892)$ bands $(792\MeV<M(\k \pi)<992\MeV)$. 
a) The scatter plot of $\k \pi$ combinations.
b) and c) are the $\k \pi$ projections fitted with a Gaussian over a linear background.
d) The $\kos \k^{\pm}$ mass spectrum; the Breit--Wigner function with the a$_0$(980) parameters
is drawn with arbitrary normalisation.}
\label{fig:daliz}
\end{figure}

To improve the statistics for the study of the intermediate decay states 
 \kstark{} and a$_0\pi$ the transverse momentum cut is released to  
 $\spt < 0.2 \GeV ^2$. This  doubles the statistics in the peak to
 $65\pm11$  events, as compared to the sample with  $\spt < 0.02 \GeV ^2$, while the contribution
of  \fb{}  is still only $\sim$10\%. To search for signals due to 
 \kstar{}(892) and a$_0$(980), the
 $\k\pi$ and $\kos\k^{\pm}$ mass spectra in the \e{} region, 
$1370\MeV<{\rm M}(\kskpi )<1560\MeV$, are investigated. The results
are shown in  Figure~\ref{fig:daliz}. A clear \kstar{}(892) signal is seen in the 
$\kos\pi^{\pm} $ and $\k^{\pm} \pi^{\mp} $ spectra. 
 Gaussian fits over  linear backgrounds give a mass and width   consistent with the \kstar{}(892).
Since the  a$_0$(980) region of the $\kos\k^{\pm}$ mass lies entirely within the \kstar{} mass bands, 
no exclusive selection of this decay channel is possible. The  $\kos\k^{\pm}$ spectrum
 (Figure~\ref{fig:daliz}d) shows no clear evidence for the presence of the  a$_0$(980). 
With the present  limited statistics no firm conclusion can be drawn concerning  the possible 
presence of the $\eta _L\ra{\rm a}_0\pi$ state in the data.  

\section{The \boldmath{\epem\ra \epem \etapipi} channel}

\subsection{ Event selection}
The $\eta$ is detected via its decay $\eta \ra \gamma \gamma$. The
events are selected by requiring two pions of opposite charge and two photons, 
identified with the criteria defined in section~\ref{sec:kkpisel}.
In addition the most energetic photon must have energy greater than  300~\MeV{} and  
the effective mass of the two photons must be inside the $\eta$ mass range, $0.47-0.62$~\GeV{}. 
Also a kinematical fit, constrained to the $\eta$ mass, is applied.

After these cuts, 6444 events are selected with \etapipi{} masses below 1750~\MeV{}. 
The $\etapipi$ mass spectrum, shown in  Figure~\ref{fig:etapipi}a, is dominated  by 
the $\eta '(958)$ resonance. A Gaussian fit gives $M=957.7 \pm 0.3$~\MeV{} 
and
$\sigma = 7.9 \pm 0.3$~\MeV{}, corresponding to the expected mass resolution in this region. 
For masses of  1280~\MeV{} and 1400~\MeV{}, the expected resolutions are  
$\sigma = 20$ \MeV{} and 24 \MeV{} respectively. 

\subsection{Two--photon width}

\begin{figure}[p]
\begin{center}
\includegraphics[width=\figwidth ]{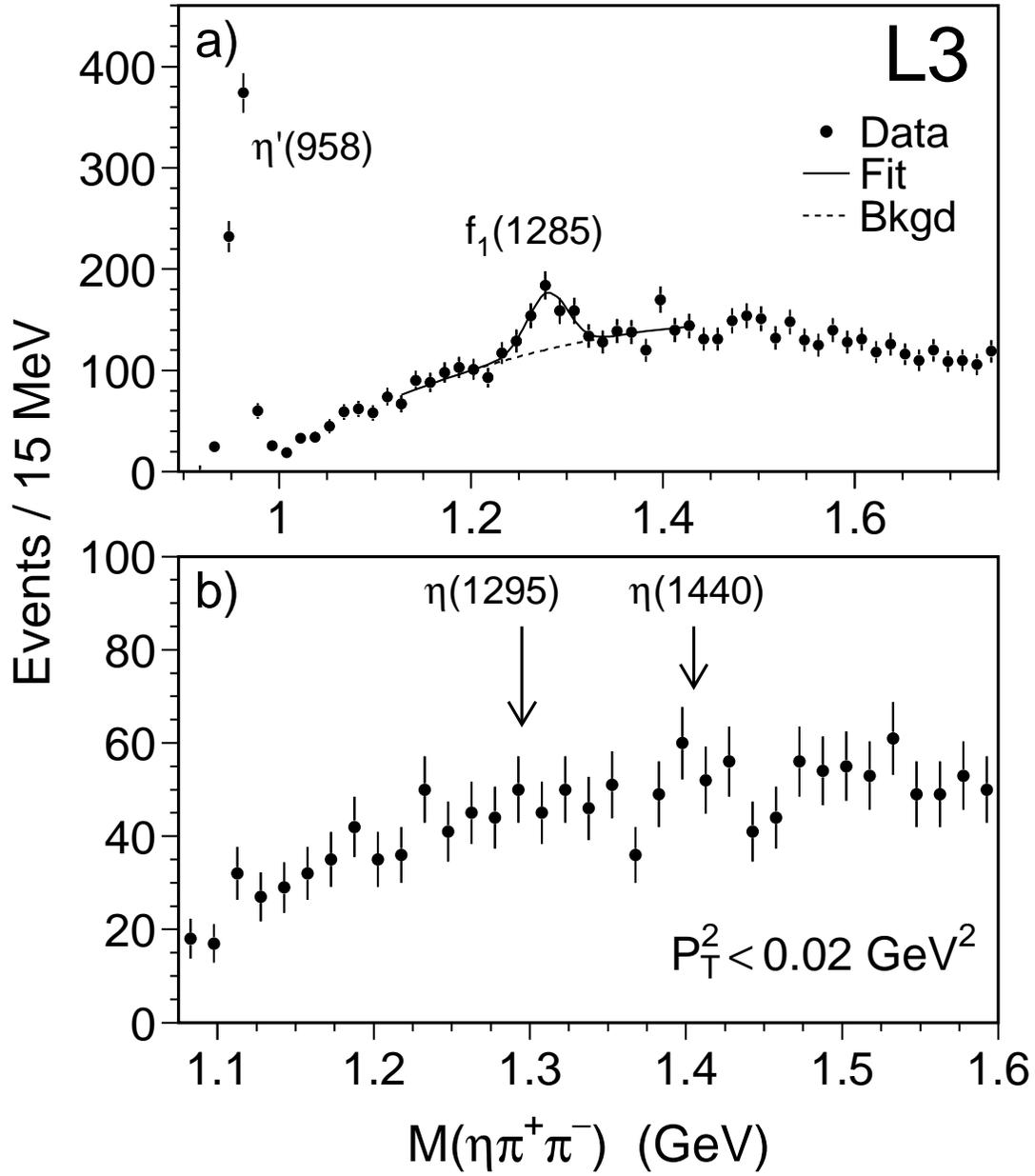}
\end{center}
\caption{The \etapipi{} mass spectra: 
a) total spectrum, the fit of a Gaussian plus polynomial background for the \fa{} region
is superimposed on the data;    
b) for $\spt < 0.02 \GeV ^2$, arrows show the location of \ea{} and \e{}.}
\label{fig:etapipi}
\end{figure}

The \etapipi{} mass spectrum for $\spt < 0.02 \GeV ^2$ is shown in Figure~\ref{fig:etapipi}b.
No peak is seen in the region 1200$-$1480~\MeV{}.
The detection efficiency is 2\% 
and includes the trigger efficiency, which varies from  60\% to 75\%. 
 Corrections for the inefficiencies due to higher level  triggers
range from 10\% to 35\%.

The absence of signals of \e{} and \ea{} allows to calculate upper limits, 
at 95\% CL on the $\eta \pi \pi$ decay.
Taking into account the branching ratio  $\BR(\eta\ra\gg)$ \cite{PDG} and 
the isospin factor $\eta\pi\pi\,/\,\eta\pip\pim $, they are: 
\begin{eqnarray}
\Ggg\Bigl(\e\Bigr)\times \BR\Bigl(\e\ra \eta \pi \pi\Bigr) & < & 95\eV \> , \nonumber \\[0.4ex]
\Ggg\Bigl(\ea\Bigr)\times \BR\Bigl(\ea\ra \eta \pi \pi\Bigr) & < & 66\eV \> .  \nonumber 
\end{eqnarray}
\noindent
They improve the limits of the Crystal Ball Collaboration \cite{crystal}.

The peak around 1285~\MeV{}, seen in Figure~\ref{fig:etapipi}a and absent in Figure~\ref{fig:etapipi}b 
at $\q2 \simeq 0$, is identified with the spin--1 state \fa{}. A Gaussian fit plus polynomial background gives
$M=1280 \pm 4$~\MeV{} and $\sigma = 21 \pm 4$~\MeV{}.

\section{Discussion}

The two--photon width of the \e{}, calculated under the assumption that it is a member 
of the first radial excitation of the pseudoscalar nonet \cite{Anisovich}, is of the order of 0.1~\keV.
The \ssbar{} content gives a negligible contribution. This is in agreement with our measurement, 
if we assume that $\BR\Bigl(\e\ra\kkbar\pi\Bigr) \sim 100 \%$. 

If the \e{} we observe is a gluon rich state, its stickiness parameter 
\cite{Chanowitz} is expected to be large.
The stickiness of a resonance R with mass $m_{\rm R}$ and two--photon width $\Gamma _{{\rm R} \ra \gg}$ 
is defined as:
$$
\frac{ |\langle {\rm R}|{\rm gg}\rangle|^2}{|\langle {\rm R}|\gg\rangle|^2}\sim
S_{\rm R} = N_l \left(\frac{m_{\rm R}}{K_{{\rm J}\to\gamma {\rm R}}}\right)^{2l+1}
\frac{\Gamma _{{\rm J}\to\gamma {\rm R}}}{\Gamma _{{\rm R} \ra \gg}} \ ,
$$
where $K_{{\rm J}\to\gamma {\rm R}}$ is the energy of the photon  in the J rest frame, 
$l$ is the orbital angular momentum of the two initial photons or gluons ($l=1$ for $0^-$), 
$\Gamma _{{\rm J}\to\gamma {\rm R}}$ is the J radiative decay width for R, 
and $N_l$ is a normalisation factor chosen to give $S_{\eta} = 1$.
Using the present measurement of $\Ggg\Bigl(\e\Bigr)\times \BR\Bigl(\e \ra \kkbar \pi\Bigr)$
and $\Gamma({\rm J}\to\gamma \e \to \gamma \kkbar \pi) = 79\pm16\eV$ \cite{PDG},
a value of the stickiness $S_{\e}=79\pm 26$ is obtained.

Another parameter, the gluiness ($G$), was introduced \cite{Close,Paar}
to quantify the ratio of the two--gluon and two--photon coupling of a particle,
it is defined as:
$$ 
G = \frac{9\,e^4_q}{2}\,\biggl(\frac{\alpha}{\alpha _s}\biggr)^2 \,
\frac{\Gamma _{{\rm R} \ra {\rm gg}}}{\Gamma _{{\rm R} \ra \gg}} \ ,
$$
where $e_q$ is the relevant quark charge, calculated assuming equal amplitudes for 
\uubar{} and \ddbar{} and zero amplitude for \ssbar{}. $\Gamma _{{\rm R} \ra {\rm gg}}$ is 
the two--gluon width of the resonance {\rm R}, 
calculated from equation (3.4) of Reference \citen{Close}.
Whereas stickiness is a relative measure, 
the gluiness is a normalised quantity and is expected to be near unity for a \qqbar{} meson.
Using the present measurement of $\Ggg\Bigl(\e\Bigr)\times \BR\Bigl(\e\ra\kkbar\pi\Bigr)$ and the average value of 
$\alpha_s(1440 \MeV) =0.369\pm0.022$  \cite{PDG}, a value $G_{\e}=41\pm 14$ is obtained. 

From the upper limit  in the \etapipi{} channel and the value
$\Gamma({\rm J}\to\gamma \e \to\gamma \etapipi )=26\pm5\eV$ \cite{PDG}  the limits 
$S_{\e} > 87$ and $G_{\e} > 45$  at $95\%$ confidence level are obtained, compatible  with the values 
obtained for the \kskpi{} decay channel\footnote{Although the coincidence can
be fortuitous if these two final states correspond to two different pseudoscalars 
$\eta _L$ and $\eta_ H$.}. 
Both the stickiness and gluiness of the \e{} point to a large gluonium content in this resonance. 
For comparison, the $\eta '$ pseudoscalar meson has 
$S_{\eta '} = 3.6 \pm 0.3$ and $G_{\eta '} = 5.2 \pm 0.8$ for $\alpha_s(958 \MeV)=0.56\pm0.07$ \cite{PDG}.

\section{Conclusions}
The pseudoscalar meson \e{} is observed for the first time in \gg{} collisions.
The two--photon width times branching ratio is measured to be
$\Ggg\Bigl(\e\Bigr)\times \BR\Bigl(\e\ra \k \bar{\k}\pi\Bigr)= 212\pm 50\stat \pm 23\sys \eV$.
Its mass and width as well as the observation of a dominant \kstar{}(892)K decay
are  compatible with the characteristics of  the  $\eta _H$.
The measured two-photon width  is consistent with the value expected for a first  radial excitation 
of the pseudoscalar nonet. 
At the same time tests designed to establish the  gluon content of a resonance 
point to a strong gluonium admixture.

No positive signal is observed for the $\eta _L$ state, either in the $\kskpi$ channel, 
where there is no clear evidence for an a$_0$(980)$\pi$ decay, or in the $\etapipi$ channel.
Upper limits for the two--photon width of the \e{} and the \ea{} in the decay channel $\etapipi$ are determined.

The high \q2{} events show clear evidence for the formation 
of the axial vector mesons \fb{} in the decay channel $\kskpi$ and 
for the formation of \fa{} in both $\kskpi$ and $\etapipi$ channels.
  
\section*{Acknowledgements}
We wish to express our gratitude to the CERN accelerator division for
the excellent performance of the LEP machine. 
We acknowledge the contributions of  the engineers and the technicians who
have participated in the construction and maintenance of this experiment.

\bibliographystyle{l3style}

%

\clearpage
\input namelist223.tex

%

\begin{table}[htp]
\begin{center}
\begin{tabular}{|l|c|c|} 
\hline
State                              & Mass (\MeV{})   & Width (\MeV{}) \\ \hline
$\eta _L \rightarrow \eta \pi \pi$ & $1405 \pm 5 $   & 56 $\pm$ \  7  \\
$\eta _L \rightarrow \kkpi$        & $1418 \pm 1 $   & 58 $\pm$ \ 4 \\
$\eta _H \rightarrow \kkpi$        & $1475 \pm 5 $   & 81 $\pm$   11 \\ \hline
\end{tabular}
\end{center}
\caption[]{Average masses and widths for the pseudoscalars in the 1440~\MeV{} region \cite{PDG}.}
\label{tab:etapar} 
\end{table}

\vspace{1cm}

\begin{table}[htp]
\begin{center}
\begin{tabular}{|c|c|c|c|c|c|c|}
\hline
$\Delta \spt \ (\GeV ^2)$ & Events & $M$ (\MeV{}) & $\sigma$ (\MeV{}) & CL (\%) & $\epsilon$ (\%)  & $\Delta \sigma$ (pb) \\ \hline
 $0 - 0.02 $   & $37\pm 9$      & $1481\pm 12$     & $48\pm \ 9$ & 89 & 1.03 $\pm$ 0.04 & 8.0 $\pm$ 2.0 \\ 
 $0.02 - 0.2$  & $28\pm 7$      & $1473\pm 11$     & $37\pm \ 8$ & 77 & 0.85 $\pm$ 0.09 & 7.4 $\pm$ 2.3 \\ 
 $0.2 - 1.$    & $29\pm 9$      & $1435\pm 10$     & $32\pm 10$  & 99 & 1.74 $\pm$ 0.14 & 3.7 $\pm$ 1.2 \\ 
  $1 - 7 $     & $21\pm 6$      & $1452\pm 11$     & $35\pm 10$  &    & 3.49 $\pm$ 0.24 & 1.4 $\pm$ 0.4 \\[-1.2ex]
               &                &                  &             & 55 &                 &               \\[-1.2ex]
 $1 - 7 $      & $10\pm 4$      & $1290\pm 12$     & $29\pm 10$  &    &   ---           &     ---       \\ \hline
\end{tabular}
\end{center}
\caption[]{Results of the Gaussian plus polynomial background fits performed
  on the mass spectra of Figure~\ref{fig:ptbins}. The number of the events in the Gaussian,
  the mass $M$ and the width $\sigma$ are listed with the error given by the fit.
  All fits reproduce the data well, as proven by the confidence level (CL) value.
  The partial efficiency $\epsilon$ and the partial cross-section $\Delta \sigma $ (errors include
  systematic uncertainties) are also given
  for each \spt{} interval.}
\label{tab:mass} 
\end{table}

\clearpage

\end{document}

%% file: namelist223.tex
\typeout{   }     
\typeout{Using author list for paper 223 -- ? }
\typeout{$Modified: Tue Sep  5 19:04:46 2000 by clare $}
\typeout{!!!!  This should only be used with document option a4p!!!!}
\typeout{   }
%
%
%
%
%
%

\newcount\tutecount  \tutecount=0
\def\tutenum#1{\global\advance\tutecount by 1 \xdef#1{\the\tutecount}}
\def\tute#1{$^{#1}$}
\tutenum\aachen            
\tutenum\nikhef            
\tutenum\mich              
\tutenum\lapp              
\tutenum\basel             
\tutenum\lsu               
\tutenum\beijing           
\tutenum\berlin            
\tutenum\bologna           
\tutenum\tata              
\tutenum\ne                
\tutenum\bucharest         
\tutenum\budapest          
\tutenum\mit               
\tutenum\debrecen          
\tutenum\florence          
\tutenum\cern              
\tutenum\wl                
\tutenum\geneva            
\tutenum\hefei             
\tutenum\seft              
\tutenum\lausanne          
\tutenum\lecce             
\tutenum\lyon              
\tutenum\madrid            
\tutenum\milan             
\tutenum\moscow            
\tutenum\naples            
\tutenum\cyprus            
\tutenum\nymegen           
\tutenum\caltech           
\tutenum\perugia           
\tutenum\cmu               
\tutenum\prince            
\tutenum\rome              
\tutenum\peters            
\tutenum\potenza           
\tutenum\salerno           
\tutenum\ucsd              
\tutenum\santiago          
\tutenum\sofia             
\tutenum\korea             
\tutenum\alabama           
\tutenum\utrecht           
\tutenum\purdue            
\tutenum\psinst            
\tutenum\zeuthen           
\tutenum\eth               
\tutenum\hamburg           
\tutenum\taiwan            
\tutenum\tsinghua          

{
\parskip=0pt
\noindent
{\bf The L3 Collaboration:}
\ifx\selectfont\undefined
 \baselineskip=10.8pt
 \baselineskip\baselinestretch\baselineskip
 \normalbaselineskip\baselineskip
 \ixpt
\else
 \fontsize{9}{10.8pt}\selectfont
\fi
\medskip
\tolerance=10000
\hbadness=5000
\raggedright
\hsize=162truemm\hoffset=0mm
\def\r{\rlap,}
\noindent

M.Acciarri\r\tute\milan\
P.Achard\r\tute\geneva\ 
O.Adriani\r\tute{\florence}\ 
M.Aguilar-Benitez\r\tute\madrid\ 
J.Alcaraz\r\tute\madrid\ 
G.Alemanni\r\tute\lausanne\
J.Allaby\r\tute\cern\
A.Aloisio\r\tute\naples\ 
M.G.Alviggi\r\tute\naples\
G.Ambrosi\r\tute\geneva\
% ...
\newpage
\begin{list}{A}{\itemsep=0pt plus 0pt minus 0pt\parsep=0pt plus 0pt minus 0pt
                \topsep=0pt plus 0pt minus 0pt}
\item[\aachen]
 I. Physikalisches Institut, RWTH, D-52056 Aachen, FRG$^{\S}$\\
 III. Physikalisches Institut, RWTH, D-52056 Aachen, FRG$^{\S}$
\item[\nikhef] National Institute for High Energy Physics, NIKHEF, 
     and University of Amsterdam, NL-1009 DB Amsterdam, The Netherlands
\item[\mich] University of Michigan, Ann Arbor, MI 48109, USA
\item[\lapp] Laboratoire d'Annecy-le-Vieux de Physique des Particules, 
     LAPP,IN2P3-CNRS, BP 110, F-74941 Annecy-le-Vieux CEDEX, France
\item[\basel] Institute of Physics, University of Basel, CH-4056 Basel,
     Switzerland
\item[\lsu] Louisiana State University, Baton Rouge, LA 70803, USA
\item[\beijing] Institute of High Energy Physics, IHEP, 
  100039 Beijing, China$^{\triangle}$ 
\item[\berlin] Humboldt University, D-10099 Berlin, FRG$^{\S}$
\item[\bologna] University of Bologna and INFN-Sezione di Bologna, 
     I-40126 Bologna, Italy
\item[\tata] Tata Institute of Fundamental Research, Bombay 400 005, India
\item[\ne] Northeastern University, Boston, MA 02115, USA
\item[\bucharest] Institute of Atomic Physics and University of Bucharest,
     R-76900 Bucharest, Romania
\item[\budapest] Central Research Institute for Physics of the 
     Hungarian Academy of Sciences, H-1525 Budapest 114, Hungary$^{\ddag}$
\item[\mit] Massachusetts Institute of Technology, Cambridge, MA 02139, USA
\item[\debrecen] KLTE-ATOMKI, H-4010 Debrecen, Hungary$^\P$
\item[\florence] INFN Sezione di Firenze and University of Florence, 
     I-50125 Florence, Italy
\item[\cern] European Laboratory for Particle Physics, CERN, 
     CH-1211 Geneva 23, Switzerland
\item[\wl] World Laboratory, FBLJA  Project, CH-1211 Geneva 23, Switzerland
\item[\geneva] University of Geneva, CH-1211 Geneva 4, Switzerland
\item[\hefei] Chinese University of Science and Technology, USTC,
      Hefei, Anhui 230 029, China$^{\triangle}$
\item[\lausanne] University of Lausanne, CH-1015 Lausanne, Switzerland
\item[\lecce] INFN-Sezione di Lecce and Universit\`a Degli Studi di Lecce,
     I-73100 Lecce, Italy
\item[\lyon] Institut de Physique Nucl\'eaire de Lyon, 
     IN2P3-CNRS,Universit\'e Claude Bernard, 
     F-69622 Villeurbanne, France
\item[\madrid] Centro de Investigaciones Energ{\'e}ticas, 
     Medioambientales y Tecnolog{\'\i}cas, CIEMAT, E-28040 Madrid,
     Spain${\flat}$ 
\item[\milan] INFN-Sezione di Milano, I-20133 Milan, Italy
\item[\moscow] Institute of Theoretical and Experimental Physics, ITEP, 
     Moscow, Russia
\item[\naples] INFN-Sezione di Napoli and University of Naples, 
     I-80125 Naples, Italy
\item[\cyprus] Department of Natural Sciences, University of Cyprus,
     Nicosia, Cyprus
\item[\nymegen] University of Nijmegen and NIKHEF, 
     NL-6525 ED Nijmegen, The Netherlands
\item[\caltech] California Institute of Technology, Pasadena, CA 91125, USA
\item[\perugia] INFN-Sezione di Perugia and Universit\`a Degli 
     Studi di Perugia, I-06100 Perugia, Italy   
\item[\cmu] Carnegie Mellon University, Pittsburgh, PA 15213, USA
\item[\prince] Princeton University, Princeton, NJ 08544, USA
\item[\rome] INFN-Sezione di Roma and University of Rome, ``La Sapienza",
     I-00185 Rome, Italy
\item[\peters] Nuclear Physics Institute, St. Petersburg, Russia
\item[\potenza] INFN-Sezione di Napoli and University of Potenza, 
     I-85100 Potenza, Italy
\item[\salerno] University and INFN, Salerno, I-84100 Salerno, Italy
\item[\ucsd] University of California, San Diego, CA 92093, USA
\item[\santiago] Dept. de Fisica de Particulas Elementales, Univ. de Santiago,
     E-15706 Santiago de Compostela, Spain
\item[\sofia] Bulgarian Academy of Sciences, Central Lab.~of 
     Mechatronics and Instrumentation, BU-1113 Sofia, Bulgaria
\item[\korea]  Laboratory of High Energy Physics, 
     Kyungpook National University, 702-701 Taegu, Republic of Korea
\item[\alabama] University of Alabama, Tuscaloosa, AL 35486, USA
\item[\utrecht] Utrecht University and NIKHEF, NL-3584 CB Utrecht, 
     The Netherlands
\item[\purdue] Purdue University, West Lafayette, IN 47907, USA
\item[\psinst] Paul Scherrer Institut, PSI, CH-5232 Villigen, Switzerland
\item[\zeuthen] DESY, D-15738 Zeuthen, 
     FRG
\item[\eth] Eidgen\"ossische Technische Hochschule, ETH Z\"urich,
     CH-8093 Z\"urich, Switzerland
\item[\hamburg] University of Hamburg, D-22761 Hamburg, FRG
\item[\taiwan] National Central University, Chung-Li, Taiwan, China
\item[\tsinghua] Department of Physics, National Tsing Hua University,
      Taiwan, China
\item[\S]  Supported by the German Bundesministerium 
        f\"ur Bildung, Wissenschaft, Forschung und Technologie
\item[\ddag] Supported by the Hungarian OTKA fund under contract
numbers T019181, F023259 and T024011.
\item[\P] Also supported by the Hungarian OTKA fund under contract
  numbers T22238 and T026178.
\item[$\flat$] Supported also by the Comisi\'on Interministerial de Ciencia y 
        Tecnolog{\'\i}a.
\item[$\sharp$] Also supported by CONICET and Universidad Nacional de La Plata,
        CC 67, 1900 La Plata, Argentina.
\item[$\diamondsuit$] Also supported by Panjab University, Chandigarh-160014, 
        India.
\item[$\triangle$] Supported by the National Natural Science
  Foundation of China.
\end{list}
}
\vfill
